\documentclass[11pt,a4paper]{article}
\usepackage[latin9]{inputenc}
\usepackage{amsmath}
\usepackage{amsthm}
\usepackage{amssymb}
\usepackage{graphicx}

\makeatletter

\usepackage{amsfonts}
\usepackage{amsthm}
\usepackage{epstopdf}

\newif\iffigs\figstrue

\makeatother

\begin{document}
\begin{titlepage} \vspace{0.3cm}
 \vspace{1cm}

\begin{center}
\textsc{\Large{}{}{}{}{}{}{}{}\ \\[0pt] \vspace{0mm}
 Instantons: thick-wall approximation }{\Large{}{}{}{}{}{}{}{}
}\\[0pt] \textsc{\Large{}{}{}{}{}{}{}{}\ \\[0.0pt] }{\Large\par}
\par\end{center}

 


\begin{center}
\vspace{35pt}
 \textsc{V. F. Mukhanov$^{~a,b}$ and A. S.
Sorin$^{~c}$}\\[15pt] 
\par\end{center}

\begin{center}
{$^{a}$ Ludwig Maxmillian University, \\[0pt] Theresienstr. 37,
80333 Munich, Germany\\[0pt] }e-mail: \textit{\small{}{}{}{}{}{}{}{}mukhanov@physik.lmu.de}{\small{}{}{}{}{}{}{}\vspace{10pt}
 }{\small\par}
\par\end{center}

\begin{center}
{$^{b}$ {\small{}{}{}{}{}{}{}{}Korea Institute for Advanced
Study\\[0pt] Seoul, 02455, Korea}}\vspace{10pt}
\par\end{center}

\begin{center}
{$^{c}$ {\small{}{}{}{}{}{}{}{}Bogoliubov Laboratory of Theoretical
Physics\\[0pt] Joint Institute for Nuclear Research \\[0pt] 141980
Dubna, Moscow Region, Russia \\[0pt] }}e-mail: \textit{\small{}{}{}{}{}{}{}{}sorin@theor.jinr.ru}{\small{}{}{}{}{}{}{}\vspace{10pt}
 }{\small\par}
\par\end{center}

\begin{center}
 
\par\end{center}

\begin{center}
\textbf{{Abstract} } 
\par\end{center}

We develop a new method for estimating the decay probability of the
false vacuum via regularized instantons. Namely, we consider the case
where the potential is either unbounded from below or the second minimum
corresponding to the true vacuum has a depth exceeding the height of
the potential barrier. In this case, the materialized bubbles dominating
the vacuum decay naturally have a thick wall and the thin-wall approximation
is not applicable. We prove that in such a case the main contribution
to the action determining the decay probability comes from the part
of the solution for which the potential term in the equation for instantons
can be neglected compared to the friction term. We show that the developed
approximation exactly reproduces the leading order results for the
few known exactly solvable potentials. The proposed method is applied
to generic scalar field potentials in an arbitrary number of dimensions.

\newpage{}

\section{Introduction}

In this work, we consider the problem of false vacuum decay for a
scalar field with potential $V\text{\ensuremath{\left(\varphi\right)}}$,
which has a local minimum normalized to zero (false
vacuum) at $\varphi_{f}<0$, separated from the unbounded part of
the potential or from the deep true vacuum by the barrier of the height
$V_{bar}$ at $\varphi=0$ (see Fig.\ref{Figure1}).

The original approach to the general problem of false vacuum decay
in $4$-dimensional scalar field theory was elaborated in \cite{Coleman,CGM}.
This approach proved to be inapplicable for a wide class of the potentials
and was recently modified in \cite{MRS1,MRS2,MS1,MS2,MS3} for a very
general situation. The key element of our modified approach is the
correct treatment of the role of quantum fluctuations, which regularize
the Euclidean classical $O(4)$-invariant singular solutions and
lead to the appearance of a whole class of new instantons, all contributing
to the vacuum decay rate. These new instantons exist even for those
potentials for which the instantons with Coleman boundary conditions
do not exist, although the false vacuum must obviously be unstable.

\vspace{0.5cm}


\begin{figure}[hbt]
\begin{centering}
\includegraphics[height=80mm]{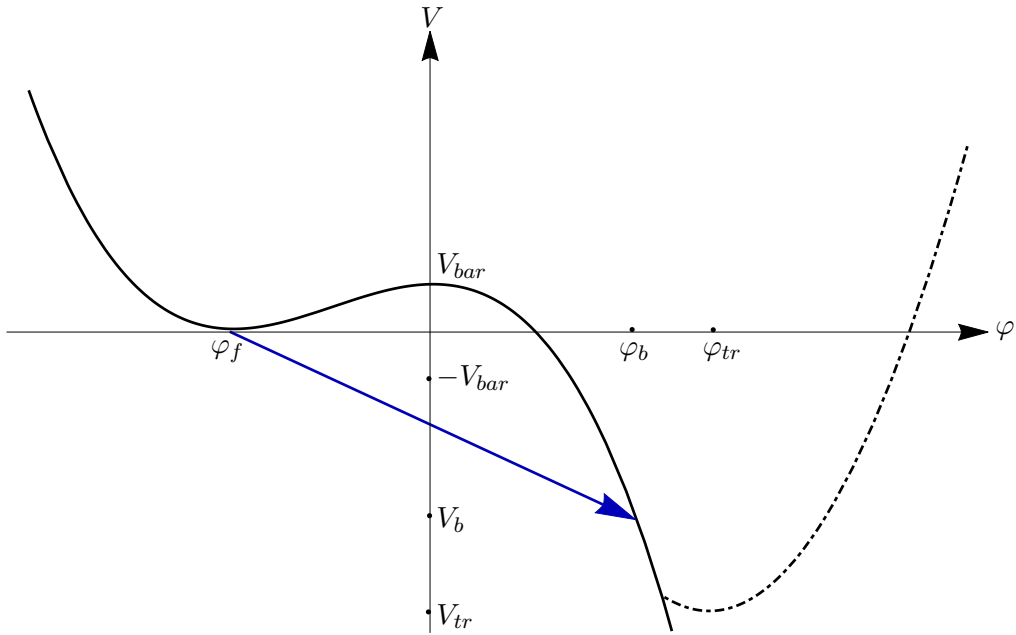} 
\par\end{centering}
\vspace*{-4.2cm}
 \hspace*{4.1cm} $\varphi_{f}$

\vspace*{-4.8cm}
 \hspace*{6.99cm}$V$

\vspace*{3.6cm}
 \hspace*{14.55cm}$\varphi$

\vspace*{0.2cm}
 \hspace*{7.2cm}$-V_{bar}$

\vspace*{1.4cm}
 \hspace*{7.2cm}$V_{b}$

\vspace*{0.8cm}
 \hspace*{7.2cm}$V_{tr}$

\vspace*{-5.12cm}
 \hspace*{7.2cm}$V_{bar}$

\vspace*{0.6cm}
 \hspace*{9.6cm}$\varphi_{b}$
 
 \vspace*{-0.48cm}
 \hspace*{10.7cm}$\varphi_{tr}$

\vspace{1.5cm}

\vspace*{3.3cm}
\caption{The potential with a metastable vacuum at $\varphi_{f}$, which must decay via instantons, regardless of whether this potential is unbounded from below or has the second true minimum represented by the dot-dashed line. For each concrete potential, there is a full spectrum of instantons. The instantons corresponding to
the deep subbarrier transition with $\left|V_{b}\right|\gg V_{bar}$ (as shown, for example, by the lower blue line) are dominated by "friction'' and cannot be described in the thin-wall approximation.}
\label{Figure1} 
\end{figure}

The aim of this work is to generalise our approach \cite{MRS1,MRS2,MS1,MS2,MS3}
to thick-wall instantons for arbitrary potentials in $D$-dimensional
spacetime. Our purpose is to derive the generic asymptotic formulas
valid in the leading order for these thick-wall bubbles.

\section{Instantons with quantum core in D dimensions}

The generalization of the basic formulas in \cite{MRS1,MRS2,MS2}
for the new instantons with quantum core in $D$-dimensional spacetime
is quite straightforward. In particular, the decay rate of the false
vacuum per unit time per unit volume must be given by 
\begin{equation}
\Gamma\simeq\varrho_{0}^{-D}\,\exp\left(-S_{I}\right)\,,\label{rate}
\end{equation}
where $\varrho_{0}$ is the typical size, i.e., ``radius'' of the
bubble and the finite on-shell instanton action is equal to
\begin{equation}
S_{I}=\frac{2\,\pi^{\frac{D}{2}}}{\Gamma\left(\frac{D}{2}\right)}\,\left(\int_{\varrho_{uv}}^{\varrho_{ir}}d\varrho\,\varrho^{D-1}\,\left(\frac{1}{2}\,\dot{\varphi}^{2}\,+\,V(\varphi)\right)+\frac{\varrho_{uv}^{D}}{D}\,V_{uv}\right).\,\label{Eeq1a}
\end{equation}
The total energy of the regularized instantons, 
\begin{equation}
{\cal V}=\frac{2\,\pi^{\frac{D-1}{2}}}{\Gamma\left(\frac{D-1}{2}\right)}\,\left(\int_{\varrho_{uv}}^{\varrho_{ir}}d\varrho\,\varrho^{D-2}\,\left(\frac{1}{2}\,\dot{\varphi}^{2}\,+\,V(\varphi)\right)+\frac{\varrho_{uv}^{D-1}}{D-1}\,V_{uv}\right)\,,\label{ins_energy}
\end{equation}
vanish at the moment of their exit from under the barrier with the
precision allowed by the time--energy uncertainty relation 
\begin{equation}
\varrho_{uv}\,{\cal V}\simeq O(1)\,.\label{unc_rel}
\end{equation}
The second terms in equations (\ref{Eeq1a}) and (\ref{ins_energy})
are the contributions from the homogeneously filled quantum core with
radius $\varrho_{uv}$ and energy--density $V_{uv}\equiv V\left(\varphi\left(\varrho_{uv}\right)\right)$,
where $\varrho_{uv}$ and $\varrho_{ir}$ are ultraviolet and infrared
cutoff scales determined by parameters of the corresponding instanton
solutions.

The instantons are the $O\left(D\right)$-invariant \textit{solutions}
of the Euclidean scalar field equation 
\begin{equation}
\frac{\partial^{2}\varphi\left(\tau,\mathbf{x}\right)}{\partial\tau^{2}}+\Delta\varphi\left(\tau,\mathbf{x}\right)-\frac{dV(\varphi)}{d\varphi}=0\,,\label{eq:first}
\end{equation}
which in this case reduces to the ordinary differential equation 
\begin{equation}
\ddot{\varphi}+\frac{D-1}{\varrho}\,\dot{\varphi}-V'=0\label{eq:2}
\end{equation}
with the boundary conditions: 
\begin{equation}
\varphi\left(\varrho\rightarrow\infty\right)=\varphi_{f}\,,\label{eq:3}
\end{equation}
\begin{equation}
\dot{\varphi}\left(\varrho=\varrho_{b}\right)=0\,.\label{eq:4}
\end{equation}
Here, $\varphi\left(\tau,\mathbf{x}\right)=\varphi\left(\varrho\right)$,
$\varrho\equiv\sqrt{\tau^{2}+\mathbf{x^{2}}}$ with Euclidean
time $\tau$, $V'\equiv dV/d\varphi$ and $\dot{\varphi}\equiv d\varphi/d\varrho$, $\varrho_{b}$ is a bouncing parameter characterizing the corresponding
instanton solution. For a given potential $V(\varphi)$, both the
ultraviolet and infrared cutoffs $\varrho_{uv}$ and $\varrho_{ir}$
can be fully expressed in terms of $\varrho_{b}$ and parameters of
the potential $V(\varphi)$, by equating the derivative of the instanton
solution $\dot{\varphi}\left(\varrho\right)$ to the derivative of
the typical amplitude of quantum fluctuations $\delta{\dot{\varphi}}\left(\varrho\right)$
in corresponding scales (see \cite{MRS1,MRS2,MS2}), that is
\begin{equation}
|\dot{\varphi}_{uv}|\equiv|\dot{\varphi}\left(\varrho_{uv}\right)|=|\delta{\dot{\varphi}}\left(\varrho_{uv}\right)|\,,\quad|\dot{\varphi}_{ir}|\equiv|\dot{\varphi}\left(\varrho_{ir}\right)|=|\delta{\dot{\varphi}}\left(\varrho_{ir}\right)|\,.\label{fluct1}
\end{equation}
In the special case $\varrho_{b}=0$ the boundary condition (\ref{eq:4})
coincides with those of Coleman \cite{Coleman} imposed to avoid a
singularity. The scaling dimension of the massless scalar field $\varphi$
in $D$ dimensional spacetime is ${cm}^{\frac{2-D}{2}}$, consequently
the typical amplitude of its quantum fluctuations in scales $\varrho$
is approximately 
\begin{equation}
|\delta\varphi_{q}(\varrho)|\simeq\sigma\,\varrho^{\frac{2-D}{2}}\,,\quad|\delta\dot{\varphi}_{q}(\varrho)|\simeq\frac{\sigma\,(D-2)}{2}\,\varrho^{-\frac{D}{2}}\,,\label{fluct2}
\end{equation}
where the parameter $\sigma$ is of order one, and the relations (\ref{fluct1})
become: 
\begin{equation}
|\dot{\varphi}_{uv}|=\frac{\sigma\,(D-2)}{2}\,\,\varrho_{uv}^{-\frac{D}{2}}\,,\quad|\dot{\varphi}_{ir}|=\frac{\sigma\,(D-2)}{2}\,\,\varrho_{ir}^{-\frac{D}{2}}\,.\label{fluct3}
\end{equation}
The classical solutions with $\varrho_{b}\neq0$ are singular at $\varrho\rightarrow0$
and have a divergent action. However, as shown in \cite{MRS1,MRS2,MS2},
these solutions for $\varrho<\varrho_{uv}$ and $\varrho>\varrho_{ir}$
are completely saturated by quantum fluctuations and therefore are
trustworthy only for $\varrho_{uv}<\varrho<\varrho_{ir}$. In particular,
the region $0<\varrho<\varrho_{uv}$, which makes an infinite contribution
to the action, must be replaced by the quantum core with radius $\varrho_{uv}$.
For these regularised instantons, the action is finite and they provide
a finite contribution to the decay rate. So instead of a single instanton
with Coleman boundary conditions (which, for example does not exist
at all for the steep unbounded potentials \cite{MS1,MS3}), we obtain
the entire spectrum of regularized instantons, all of which contribute
to the false vacuum decay.

Using the identities: 
\begin{eqnarray}
\varrho^{D-2}\,\left(\frac{1}{2}\,{\dot{\varphi}}^{2}+V(\varphi)\right)&\equiv&-\frac{d}{d\varrho}\,\left(\frac{\varrho^{D-1}}{D-1}\,\left(\frac{1}{2}\,{\dot{\varphi}}^{2}-V(\varphi)\right)\right)\nonumber\\
&+&\frac{\varrho^{D-1}}{D-1}\,{\dot{\varphi}}\,\left(\ddot{\varphi}(\varrho)+\frac{D-1}{\varrho}\,\dot{\varphi}(\varrho)-V'(\varphi)\right)\label{A_1}
\end{eqnarray}
and 
\begin{eqnarray}
\varrho^{D-1}\,\left(\frac{1}{2}\,{\dot{\varphi}}^{2}+V(\varphi)\right) & \equiv & \frac{\varrho^{D-1}}{D}\,{\dot{\varphi}}^{2}-\frac{d}{d\varrho}\,\left(\frac{\varrho^{D}}{D}\,\left(\frac{1}{2}\,{\dot{\varphi}}^{2}-V(\varphi)\right)\right)\nonumber \\
 & + & \frac{\varrho^{D}}{D}\,{\dot{\varphi}}\,\left(\ddot{\varphi}(\varrho)+\frac{D-1}{\varrho}\,\dot{\varphi}(\varrho)-V'(\varphi)\right),\label{A_2}
\end{eqnarray}
where the last terms on the right-hand sides vanish on-shell (see
(\ref{eq:2})), we can simplify the expressions for the total energy
${\cal V}$ (\ref{ins_energy}) and for the action $S_{I}$ (\ref{Eeq1a})
as: 
\begin{eqnarray}
{\cal V}=\frac{\pi^{\frac{D-1}{2}}}{\Gamma\left(\frac{D+1}{2}\right)}\,\left(\frac{\varrho_{uv}^{D-1}}{2}\,{\dot{\varphi}_{uv}}^{2}-\varrho_{ir}^{D-1}\,\left(\frac{1}{2}\,{\dot{\varphi}}_{ir}^{2}-V(\varphi_{ir})\right)\right)\approx\frac{\sigma^{2}\,\pi^{\frac{D-1}{2}}}{2\,\Gamma\left(\frac{D+1}{2}\right)}\,\left(1-\frac{\varrho_{uv}}{\varrho_{ir}}\right)\,\frac{1}{\varrho_{uv}}\,\label{A_1a}
\end{eqnarray}
and 
\begin{eqnarray}
S_{I} & = & \frac{\pi^{\frac{D}{2}}}{\Gamma\left(\frac{D+2}{2}\right)}\,\Bigg(\int_{\varrho_{uv}}^{\varrho_{ir}}d\varrho\,\varrho^{D-1}\,{\dot{\varphi}}^{2}+\frac{\varrho_{uv}^{D}}{2}\,{\dot{\varphi}_{uv}}^{2}-\varrho_{ir}^{D}\,\left(\frac{1}{2}\,{\dot{\varphi}}_{ir}^{2}-V(\varphi_{ir})\right)\Bigg)\nonumber \\
 & \approx & \frac{\pi^{\frac{D}{2}}}{\Gamma\left(\frac{D+2}{2}\right)}\,\int_{\varrho_{uv}}^{\varrho_{ir}}d\varrho\,\varrho^{D-1}\,{\dot{\varphi}}^{2}\,,\label{A_2a}
\end{eqnarray}
where we have also used relations (\ref{fluct3}) and assumed that
$\varrho_{ir}^{D-1}\,V(\varphi_{ir})\approx0$ and $\varrho_{ir}^{D}\,V(\varphi_{ir})\simeq O(1)$
as $\varrho_{ir}\rightarrow\infty$, as well as we have neglected
the term $\varrho_{ir}^{D}\,V(\varphi_{ir})$ of order one in the
action $S_{I}$, because the semiclassical approximation is valid
anyway only when $S_{I}\gg1$. Moreover, if we consider that 
\begin{equation}
\int_{\varrho_{b}}^{\varrho_{uv}}d\varrho\,\varrho^{D-1}\,{\dot{\varphi}}^{2}<\frac{1}{D}\,\dot{\varphi}_{uv}^{2}\,\varrho_{uv}^{D}\simeq O(1)\,,\quad\int_{\varrho_{ir}}^{\infty}d\varrho\,\varrho^{D-1}\,{\dot{\varphi}}^{2}\simeq O(1)\,,\label{29}
\end{equation}
then finally the instanton action can be rewritten in a very simple
way 
\begin{equation}
S_{I}\approx\frac{\pi^{\frac{D}{2}}}{\Gamma\left(\frac{D+2}{2}\right)}\,\int_{\varrho_{b}}^{\infty}d\varrho\,\varrho^{D-1}\,{\dot{\varphi}}^{2}=\frac{\pi^{\frac{D}{2}}}{\Gamma\left(\frac{D+2}{2}\right)}\,\int_{\varphi_{b}}^{\varphi_{f}}d\varphi\varrho^{D-1}\dot{\varphi}\,,\label{30}
\end{equation}
where $\varphi_b\equiv \varphi(\varrho_b)$ and both the ultraviolet and infrared cutoff scales do not enter
at all. Since $\varrho_{uv}<\varrho_{ir}$, it follows from (\ref{A_1a})
that the total energy ${\cal V}$ actually vanishes with the required
precision allowed by the time-energy uncertainty relation (\ref{unc_rel}).
This satisfy the necessary condition for the bubble with quantum core
to emerge from under the barrier at $\varrho=\varrho_{uv}$. The bubble
materializes and expands, filling the Minkowski space with a new phase.

\section{Friction dominated instantons}

To calculate the false vacuum decay rate, we need first to solve the
nonlinear equation (\ref{eq:2}) with boundary conditions (\ref{eq:3}--\ref{eq:4})
and then determine the value of the instanton action $S_{I}$ (\ref{30})
for this solution. This equation can be solved exactly only for some
very special potentials.

Coleman \cite{Coleman} has developed the thin-wall approximation
method, which allows us to estimate the contribution of leading order
to the action when the depth of the tunnelling $\left|V_{b}\right|\equiv \left|V(\varphi_b)\right|$
(the magnitude of the potential at the center of the emergent bubble)
is much smaller than the height of the potential barrier $V_{bar}$,
i.e., $\left|V_{b}\right|\ll V_{bar}$. In this case, the friction
term in equation (\ref{eq:2}) can be initially neglected and then
considered perturbatively.

In the opposite case of deep tunnelling, for $\left|V_{b}\right|\gg V_{bar}$,
as shown in \cite{MS2}, the friction term dominates over major
part of the instanton, which gives the main contribution to the action.
Such instantons determine the decay rate for the unbounded potentials
or for the potentials with the deep true vacuum. The main aim of this
work is to develop a general method for calculating the decay rate
in the case when it is determined by the friction dominated instantons,
such as for the unbounded potentials shown in Fig.\ref{Figure1}.

When the friction term in equation (\ref{eq:2}) dominates over the
potential term, i.e., 
\begin{equation}
\Big|\frac{D-1}{\varrho}\,\dot{\varphi}\Big|\gg V',\label{eq:31}
\end{equation}
this equation simplifies to

\begin{eqnarray}
\ddot{\varphi}+\frac{D-1}{\varrho}\,\dot{\varphi}\simeq0\label{F2}
\end{eqnarray}
and has an obvious solution 
\begin{eqnarray}
{\varphi}(\varrho)=\varphi_{f}+\frac{E}{(D-2)^{2}\,|\varphi_{f}|\,\varrho^{D-2}}\,\label{F4}
\end{eqnarray}
which satisfies the boundary condition (\ref{eq:3}). We have introduced
here the constant of integration $E$ which in principle can be expressed
in terms of $\varphi_b\equiv \varphi(\varrho_b)$, where $\varrho_b$ enters the second boundary condition
(\ref{eq:4}). Therefore, $E$ can be used instead of $\varphi_{b}$
to parametrize the regularized friction dominated instantons. One
can easily verify that the friction dominated solution (\ref{F4})
satisfies the following useful relation 
\begin{equation}
E=\varrho^{D}\,\left(\dot{\varphi}^{2}+\frac{D-2}{\varrho}\,\varphi\,\dot{\varphi}\right)\,.\label{eq:32}
\end{equation}
Estimating the expression on the right-hand side of this equality
for quantum fluctuations (\ref{fluct2}) we get 
\begin{equation}
\varrho^{D}\,\left(|\delta\dot{\varphi}_{q}(\varrho)|^{2}+\frac{D-2}{\varrho}|\delta\varphi_{q}(\varrho)||\delta\dot{\varphi}_{q}(\varrho)\right)\simeq\frac{3(D-1)^{2}\,\sigma^{2}}{4}\sim O(1)\,,\label{eq:33}
\end{equation}
which is independent on the scale $\varrho$. From this we conclude
that $E$ must be much larger than unity, i.e., 
\begin{eqnarray}
E\gg1\,,\label{EEEEE-1}
\end{eqnarray}
otherwise the quantum fluctuations would dominate everywhere over
the friction part of the classical solution (mean field) and therefore
its contribution to the action cannot be trusted at all.

An important property of the solution (\ref{F4}) is that it can be
inverted and one can express $\varrho$ in terms of $\varphi$: 
\begin{eqnarray}
\varrho(\varphi)=\left(\frac{E}{(D-2)^{2}\,|\varphi_{f}|\,(\varphi-\varphi_{f})}\right)^{\frac{1}{D-2}}\,.\label{F4a}
\end{eqnarray}
As a result, the equation (\ref{F2}) can be rewritten in the following
autonomous (no explicit $\varrho$ dependence) form 
\begin{eqnarray}
{\ddot{\varphi}(\varrho)}+V'_{fr}({\varphi})=0\,,\label{F2_aut1}
\end{eqnarray}
where 
\begin{eqnarray}
V_{fr}({\varphi})\equiv-\frac{1}{2}\,(D-2)^{\frac{2D}{D-2}}\,\left(\frac{|\varphi_{f}|}{E}\right)^{\frac{2}{D-2}}\,\left(\varphi-\varphi_{f}\right)^{\frac{2(D-1)}{D-2}}\leq0\label{F2_aut2}
\end{eqnarray}
is the ``friction potential''. The equation (\ref{F2_aut1}) must
be supplemented by the boundary condition at $\varrho\rightarrow\infty$
(\ref{eq:3}) and the condition that its first integral vanishes,
i.e., 
\begin{eqnarray}
\frac{1}{2}\,{\dot{\varphi}(\varrho)}^{2}+V_{fr}({\varphi})=0\,.\label{F2_aut4}
\end{eqnarray}
In this case, its solution corresponds exactly to (\ref{F4}).

This consideration suggests that for the cases where the friction
dominated part of the instanton is the main contributor to the action,
one can replace the exact equation (\ref{eq:2}) with autonomous equation
\begin{equation}
{\ddot{\varphi}}+U'_{eff}({\varphi})=0\,,\label{eq:34}
\end{equation}
where 
\begin{equation}
U_{eff}=V_{fr}-V\,.\label{eq:35}
\end{equation}
The first boundary condition for this equation must be as before (\ref{eq:3})
and the second one follows from equating the first integral of this
equation to zero by analogy with (\ref{F2_aut4}), that is 
\begin{equation}
\frac{1}{2}\,{\dot{\varphi}}^{2}+U_{eff}=0\,.\label{eq:36}
\end{equation}
Let us now determine under which conditions the non-autonomous equation
(\ref{eq:2}) can be well approximated by its nontrivial substitution
by the autonomous equation (\ref{eq:34}). As follows from (\ref{eq:36}),
the value of the scalar field at which its velocity vanishes must
satisfy 
\begin{equation}
U_{eff}(\varphi)=0\,.\label{eq:37}
\end{equation}
One of the solutions of this equation is obvious, namely, the location
of the false vacuum, i.e., $\varphi=\varphi_{f}.$ The other solution
determines the bouncing value $\varphi_{b}$ for the case when the
main part of the instanton is dominated by friction, i.e., satisfies
the equation (\ref{F2}). It is clear that $\varphi_{b}$ must lie
to the right of the maximum of the potential, i.e., $\varphi_{b}>0$.
If we assume that $\dot{\varphi}$ at the location of the maximum
of the potential $V(\varphi=0)=V_{bar}$ is determined by the friction
term, then it follows from (\ref{eq:36}) that the friction dominated
instanton must satisfy the following necessary condition 
\begin{equation}
\left|V_{fr}(\varphi=0)\right|\gg V_{bar}\,,\label{eq:38}
\end{equation}
and using (\ref{F2_aut2}) we find that for these instantons
\begin{equation}
1\ll E\ll(D-2)^{D}|\varphi_{f}|^{D}\,\left(2V_{bar}\right)^{\frac{2-D}{2}}.\label{eq:39}
\end{equation}
As follows from (\ref{eq:37}), at the bounce point $\varphi_{b}$
the depth of the tunnelling is 
\begin{equation}
V_{b}\equiv V(\varphi_{b})\simeq V_{fr}(\varphi_{b})\,.\label{eq:40}
\end{equation}
Since $\left|V_{fr}(\varphi_{b})\right|\geq\left|V_{fr}(0)\right|$,
the condition (\ref{eq:38}) can be rewritten in a slightly stronger
form, namely, 
\begin{equation}
\left|V_{b}\right|\gg V_{bar}\,,\label{eq:41}
\end{equation}
i.e., when the tunnelling depth is much larger than the height of the
potential barrier, the instantons are mainly dominated by the friction
term. This is converse to the condition for the applicability of thin-wall
approximation, that is why we alternatively refer to these instantons
as thick-wall instantons. It is interesting to note that in equation
(\ref{eq:2}) the friction term becomes comparable to the potential
term at $\varphi_{m}$, which satisfies 
\begin{equation}
\left|V'(\varphi_{m})\right|\simeq\left|V'_{fr}(\varphi_{m})\right|.\label{eq:42}
\end{equation}
Comparing this with (\ref{eq:40}), we see that for the thick-wall
instantons the potential term $V'$ becomes important only near the
bounce point, i.e., at $(\varphi_{b}-\varphi_{m})/\varphi_{m}\sim O(1)$.

\section{The decay rate}

To calculate the false vacuum decay rate, we need to find the typical
size of the bubble and the action (\ref{30}) for the corresponding
thick-wall instantons. To calculate the action, we first rewrite the
equation (\ref{eq:2}) into the following form 
\begin{equation}
\frac{d}{d\varphi}\left(\varrho^{D-1}\dot{\varphi}\right)=\frac{V'\varrho^{D-1}}{\dot{\varphi}}\,.\label{eq:44}
\end{equation}
After integration with taking into account the boundary condition
at $\varphi\rightarrow\varphi_{f}$ we get 
\begin{equation}
\varrho^{D-1}\dot{\varphi}=-\frac{E}{(D-2)\left|\varphi_{f}\right|}+\int_{\varphi_{f}}^{\varphi}d\varphi\frac{V'\varrho^{D-1}}{\dot{\varphi}}\,.\label{eq:45}
\end{equation}
Before we use this expression in the action (\ref{30}), we estimate
the integral on the right-hand side of this equation. For $\varphi<\varphi_{m}$
the friction term dominates and therefore we can substitute into the
integral 
\begin{equation}
\frac{D-1}{\varrho}\,\dot{\varphi}\simeq V'_{fr},\quad\varrho^{D-2}\simeq\frac{E}{(D-2)^{2}\,|\varphi_{f}|\,(\varphi-\varphi_{f})}\label{eq:46}
\end{equation}
and obtain 
\begin{equation}
\int_{\varphi_{f}}^{\varphi}d\varphi\frac{V'\varrho^{D-1}}{\dot{\varphi}}\simeq\frac{E}{2(D-2)\left|\varphi_{f}\right|}\int_{\varphi_{f}}^{\varphi}d\varphi\frac{V'}{V_{fr}}\simeq\frac{E}{2(D-2)\left|\varphi_{f}\right|}\left(\frac{V(\varphi)}{V_{fr}}+...\right),\label{eq:47}
\end{equation}
where we have used the explicit expression (\ref{F2_aut2}) for $V_{fr}$.\footnote{To simplify the consideration, we assume that near the false vacuum
$V\propto(\varphi-\varphi_{f})^{\alpha},$ where $\alpha>2(D-1)/(D-2).$ } It is clear that this integral term is small compared to the first
term on the right-hand side of (\ref{eq:45}) up to almost the bounce
and thus
\begin{equation}
\varrho^{D-1}\dot{\varphi}\simeq-\frac{E}{(D-2)\left|\varphi_{f}\right|}\label{eq:47a}
\end{equation}
in the interval $\varphi_{f}<\varphi<\varphi_{m}$.

To find the behavior of $\varrho^{D-1}\dot{\varphi}$ near the bounce,
where the potential dominates over the friction potential, we consider
that this happens in the interval $\varphi_{m}<\varphi<\varphi_{b}$,
where $(\varphi_{b}-\varphi_{m})/\varphi_{m}\sim O(1)$. In this case,
$V'(\varphi)$ can be approximated by its value at the bounce $V'_{b}\equiv V'(\varphi_{b})$,
and taking into account that $\dot{\varphi}(\varrho_{b})=0$, we obtain
from (\ref{eq:44}) 
\begin{equation}
\varrho^{D-1}\dot{\varphi}\simeq\frac{V'_{b}}{D}\left(\varrho^{D}-\varrho_{b}^{D}\right).\label{eq:48}
\end{equation}
This equation can be further integrated further to give 
\begin{equation}
\varphi=\varphi_{b}+\frac{V'_{b}}{2D}\varrho^{2}\left[1-\frac{D}{D-2}\left(\frac{\varrho_{b}}{\varrho}\right)^{2}+\frac{2}{D-2}\left(\frac{\varrho_{b}}{\varrho}\right)^{D}\right]\,.\label{eq:49}
\end{equation}
For $\varrho>\varrho_{b}$ we can neglect the last two terms inside
the parentheses and express $\varrho$ in terms of $\left(\varphi_{b}-\varphi\right)$.
Then equation (\ref{eq:48}) can be rewritten as 
\begin{equation}
\varrho^{D-1}\dot{\varphi}\simeq-2\left(\frac{2D}{\left|V'_{b}\right|}\right)^{\frac{D-2}{2}}\left(\varphi_{b}-\varphi\right)^{\frac{D}{2}}\label{eq:50}
\end{equation}
for the interval $\varphi_{m}<\varphi<\varphi_{b}.$ The contribution
of this term to the action is about 
\begin{equation}
S_{pot}\sim\int_{\varphi_{b}}^{\varphi_{m}}d\varphi\varrho^{D-1}\dot{\varphi}\sim\left|V'_{b}\right|^{\frac{2-D}{2}}\left(\varphi_{b}-\varphi_{m}\right)^{\frac{2+D}{2}}\sim\left|V_{b}\right|^{\frac{2-D}{2}}\varphi_{b}^{D}\,,\label{eq:51}
\end{equation}
where we skipped all numerical coefficients of order one and made
the following rough estimates $\varphi_{m}\sim\varphi_{b}$ and $V'_{b}\sim V_{b}/\varphi_{b}$.
This contribution never exceeds the contribution of the friction dominated
part of the instanton, which, as can be seen using (\ref{eq:47a}) is
of order 
\begin{equation}
S_{fr}\sim\int_{\varphi_{m}}^{\varphi_{f}}d\varphi\varrho^{D-1}\dot{\varphi}\sim\frac{E}{\left|\varphi_{f}\right|}\left(\varphi_{m}-\varphi_{f}\right)\sim\frac{E\left(\varphi_{b}-\varphi_{f}\right)}{\left|\varphi_{f}\right|}\,.\label{eq:52}
\end{equation}
Then from (\ref{eq:51}) and (\ref{eq:52}) it follows
\begin{equation}
\frac{S_{pot}}{S_{fr}}\sim\left(\frac{\varphi_{b}}{\varphi_{b}-\varphi_{f}}\right)^{D},\label{eq:53}
\end{equation}
where we have taken into account that $V_{b}=V_{fr}\left(\varphi_{b}\right)$
and used the expression (\ref{F2_aut2}) for the friction potential.
Thus, if $\varphi_{b}$ determined from the equation 
\begin{equation}
V\left(\varphi_{b}\right)=V_{fr}\left(\varphi_{b}\right)\label{eq:54}
\end{equation}
is much smaller than $\left|\varphi_{f}\right|$, then $S_{pot}\ll S_{fr}$, while for $\varphi_{b}\gg\left|\varphi_{f}\right|$ both terms give
the contribution of same order, i.e., $S_{pot}\sim S_{fr}$. Therefore
\begin{equation}
S=\frac{\alpha\,\pi^{\frac{D}{2}}E\left(\varphi_{b}-\varphi_{f}\right)}{\Gamma\left(\frac{D+2}{2}\right)(D-2)\left|\varphi_{f}\right|}=\frac{\alpha\,(D-2)^{D-1}\pi^{\frac{D}{2}}}{\Gamma\left(\frac{D+2}{2}\right)}\frac{\left(\varphi_{b}+\left|\varphi_{f}\right|\right)^{D}}{\left|2V_{b}\right|^{\frac{D-2}{2}}}\,,\label{eq:55}
\end{equation}
where the numerical coefficient $\alpha\rightarrow1$ for $\varphi_{b}\ll\left|\varphi_{f}\right|$
and it is of order one, i.e., $\alpha\sim O(1)$ if $\varphi_{b}\gg\left|\varphi_{f}\right|$.
Note that in the case $\varphi_{b}\ll\left|\varphi_{f}\right|$ the
numerical coefficient in the action is exact in the leading order,
while for $\varphi_{b}\gg\left|\varphi_{f}\right|$ the action is
estimated even in the leading order with an accuracy up to a numerical
factor of order one, the concrete value of which depends on the potential
$V(\varphi)$.

To obtain the expression for the typical ``size'' of the thick-wall
bubble, we note that the friction term still dominates when the field
crosses $\varphi=0$ and if we define the ``size of the bubble''
as $\varrho_{0}=\varrho(\varphi=0)$ we obtain from (\ref{F4a}) 
\begin{equation}
\varrho_{0}=\left(\frac{E}{(D-2)^{2}\,\varphi{}_{f}^{2}}\right)^{\frac{1}{D-2}}=\frac{\left(D-2\right)}{\sqrt{\left|2V_{b}\right|}}\left(1+\frac{\varphi_{b}}{\left|\varphi_{f}\right|}\right)^{\frac{D-1}{D-2}}\left|\varphi_{f}\right|\,.\label{eq:56}
\end{equation}
To express $E$ in the formulas (\ref{eq:55}) and (\ref{eq:56})
in terms of $\varphi_{b}$ and $V_{b}$, we considered that $V_{b}=V_{fr}\left(\varphi_{b}\right)$
and used the expression (\ref{F2_aut2}) for $V_{fr}$. Thus, one
can use the value of the scalar field at the bounce (in the ``center
of the bubble'') $\varphi_{b}$ to parametrize the regularized instantons.
The condition (\ref{eq:39}) under which the thick-wall approximation
is applicable can be rewritten as 
\begin{eqnarray}
\left(1+\frac{\varphi_{b}}{\left|\varphi_{f}\right|}\right)^{\frac{2(D-1)}{D-2}}V_{bar}\ll\left|V_{b}\right|\ll\frac{1}{2}\left(D-2\right)^{\frac{2D}{D-2}}\left(1+\frac{\varphi_{b}}{\left|\varphi_{f}\right|}\right)^{\frac{2(D-1)}{D-2}}\left|\varphi_{f}\right|^{\frac{2D}{D-2}}.\label{eq:57}
\end{eqnarray}
For the case when the potential has the second true minimum of depth
$\left|V_{tr}\right|$ in the interval (\ref{eq:57}), the largest
contribution to the decay rate comes from the instantons with $\left|V_{b}\right|\simeq\left|V_{tr}\right|$.
Substituting (\ref{eq:55}) and (\ref{eq:56}) into (\ref{rate}),
we obtain the contribution of the instantons parametrized by $\varphi_{b}$
to the total decay rate of the false vacuum. Note that even if the
potential is unbounded the results are only credible if $E\gg 1.$ Therefore,
to estimate the maximal depth of the potential to which the field
can tunnel, we must first determine the maximum value of $\varphi_{b}^{max}$
by equating $V(\varphi_{b})$ with $V_{fr}\left(\varphi_{b}\right)$
for $E\simeq1$, i.e., 
\begin{equation}
V(\varphi_{b}^{max})\simeq-\frac{1}{2}\left(D-2\right)^{\frac{2D}{D-2}}\left(1+\frac{\varphi_{b}^{max}}{\left|\varphi_{f}\right|}\right)^{\frac{2(D-1)}{D-2}}\left|\varphi_{f}\right|^{\frac{2D}{D-2}},\label{eq:57a}
\end{equation}
solve this equation for $\varphi_{b}^{max}$ and then determine the
maximal possible penetration depth. For example, if the potential
has a second very deep true minimum at $\varphi_{tr}$, at which $\left|V\left(\varphi_{tr}\right)\right|>V\left(\varphi_{b}^{max}\right)$
and the Coleman instanton that connects $\varphi_{f}$ with the neighbourhood
of $\varphi_{tr}$ exists, this instanton is nevertheless not trustworthy
because it is below the level of quantum fluctuations. In this case,
the main contribution to the decay rate comes from the instantons
describing tunnelling from $\varphi_{f}$ to $\varphi_{b}^{max}$.
A particular example of such a situation was presented in \cite{MRS1}
for an exactly solvable linear potential.

\section{Example}

To illustrate how the above approximation works, and to compare our
approximate formulas with the exact results, we consider the case
of an exactly solvable quartic potential in $D=4$ dimensions, which
we studied in \cite{MRS2}:

\begin{equation}
V\left(\varphi\right)=\left\{ \begin{array}{cc}
\frac{\lambda_{+}}{4}\left(\varphi-\varphi_{f}\right)^{4} & \text{for }\varphi\leq\beta\,\varphi_{f},\\
\\
-\frac{\lambda_{-}}{4}\,\varphi^{4}+V_{bar} & \text{for }\varphi\geq\beta\,\varphi_{f}\,,
\end{array}\right.\label{AS_5ab_S0_4}
\end{equation}
where $0<\beta<1$ and the the height of the barrier $V_{bar}$ are
expressed in terms of the positive coupling constants $\lambda_{+}$
and $\lambda_{-}$ as 
\begin{equation}
\beta\equiv\frac{\lambda_{+}^{\frac{1}{3}}}{\lambda_{+}^{\frac{1}{3}}+\lambda_{-}^{\frac{1}{3}}}\,,\quad V_{bar}\equiv\frac{\lambda_{-}}{4}\,\beta^{3}\,\varphi_{f}^{4}\,.\label{60_4}
\end{equation}
This potential is composed of two power-law potentials glued at $\varphi=\beta\,\varphi_{f}<0$,
so that both the potential and its first derivative are continuous
at this point. Taking into account that the thick-wall approximation
is valid only when $\left|V_{b}\right|\gg V_{bar}$, we can neglect
$V_{bar}$ in (\ref{AS_5ab_S0_4}) and use 
\begin{equation}
V_{b}\simeq-\frac{\lambda_{-}}{4}\varphi_{b}^{4}\label{eq:58}
\end{equation}
in (\ref{eq:55}) and (\ref{eq:56}) for $D=4$: 
\begin{align}
S & =\frac{8\pi^{2}\alpha}{\lambda_{-}}\left(1+\frac{\left|\varphi_{f}\right|}{\varphi_{b}}\right)^{4},\label{eq:59}\\
 & \varrho_{0}=\sqrt{\frac{8}{\lambda_{-}}}\left(1+\frac{\varphi_{b}}{\left|\varphi_{f}\right|}\right)^{3/2}\frac{\left|\varphi_{f}\right|}{\varphi_{b}^{2}}\,.\label{eq:60}
\end{align}
According to (\ref{eq:57}) these expressions are valid only when
$\varphi_{b}$ satisfies inequality
\begin{equation}
\left(1+\frac{\varphi_{b}}{\left|\varphi_{f}\right|}\right)^{3}V_{bar}\ll\frac{1}{4}\lambda_{-}\varphi_{b}^{4}\ll8\left(1+\frac{\varphi_{b}}{\left|\varphi_{f}\right|}\right)^{3}\left|\varphi_{f}\right|^{4},\label{eq:61}
\end{equation}
and therefore the corresponding instanton is friction dominated. To
compare the above formulas with those in \cite{MRS2}, presented
in terms of $E$, we need to express $\varphi_{b}$ in terms of $E.$
The corresponding relation follows directly from the equation $V_{b}=V_{fr}\left(\varphi_{b}\right)$
and in our particular example becomes 
\begin{equation}
\left(1+\frac{\left|\varphi_{f}\right|}{\varphi_{b}}\right)^{3}\frac{\left|\varphi_{f}\right|}{\varphi_{b}}=\frac{\lambda_{-}E}{32}\,.\label{eq:63}
\end{equation}
We consider separately two cases where the coupling constant $\lambda_{-}$
is either much smaller or much larger than $\lambda_{+}$.

For the potential that is very flat near its maximum, i.e., $\lambda_{-}\ll\lambda_{+}$,
the height of the potential barrier is $V_{bar}\simeq\frac{1}{4}\lambda_{-}\varphi_{f}^{4}$
and as follows from (\ref{eq:61}) 
\begin{equation}
\left|\varphi_{f}\right|\ll\varphi_{b}\ll\frac{32}{\lambda_{-}}\left|\varphi_{f}\right|\,.\label{eq:64}
\end{equation}
In this case $\varphi_{b}$ is always larger than $\left|\varphi_{f}\right|$
and it follows from (\ref{eq:63}) that
\begin{equation}
\varphi_{b}\simeq\frac{32}{\lambda_{-}E}\left|\varphi_{f}\right|\label{eq:65}
\end{equation}
for $1\ll E\ll32/\lambda_{-}$. The formulas (\ref{eq:59}) and (\ref{eq:60})
simplify to 
\begin{equation}
S\simeq\frac{8\,\pi^{2}\alpha}{\lambda_{-}},\qquad\varrho_{0}\simeq\sqrt{\frac{E}{4}}\frac{1}{\left|\varphi_{f}\right|}\,.\label{eq:66}
\end{equation}
Comparing these results to those from \cite{MRS2}, we see that $\alpha=1/3.$
All other numerical coefficients in the expressions for $\varphi_{b}$
and $\varrho_{0}$ are exact in the leading order.

For the potential that drops very rapidly after reaching its maximum,
i.e., $\lambda_{-}\gg\lambda_{+}$ the interval of $\varphi_{b}$
for which our thick-wall approximation holds is larger. Considering
that in this case $V_{bar}\simeq\frac{1}{4}\lambda_{+}\varphi_{f}^{4}$,
we find that the inequality (\ref{eq:61}) in this case reduces
to 
\begin{equation}
\left(\frac{\lambda_{+}}{\lambda_{-}}\right)^{1/4}\left|\varphi_{f}\right|\ll\varphi_{b}\ll\frac{32}{\lambda_{-}}\left|\varphi_{f}\right|\,.\label{eq:67}
\end{equation}
In the range (\ref{eq:64}), the expressions (\ref{eq:65}) and (\ref{eq:66})
remain unchanged. For 
\begin{equation}
\left(\frac{\lambda_{+}}{\lambda_{-}}\right)^{1/4}\left|\varphi_{f}\right|\ll\varphi_{b}\ll\left|\varphi_{f}\right|,\label{eq:68}
\end{equation}
the solution of the equation (\ref{eq:63}) is 
\begin{equation}
\varphi_{b}\simeq\left(\frac{32}{\lambda_{-}E}\right)^{1/4}\left|\varphi_{f}\right|,\label{eq:69}
\end{equation}
and it holds for $32/\lambda_{-}\ll E\ll32/\lambda_{+}.$ Since for
$\varphi_{b}\ll\left|\varphi_{f}\right|$ the value of $\alpha$ tends
to one, the expressions (\ref{eq:59}) and (\ref{eq:60}) simplify
to 
\begin{equation}
S=\frac{\pi^{2}}{4}E,\qquad\varrho_{0}\simeq\sqrt{\frac{E}{4}}\frac{1}{\left|\varphi_{f}\right|}\label{eq:70}
\end{equation}
in full agreement with the results derived in \cite{MRS2} from exact
solutions. Note, that the maximum depth of tunnelling via classical
instantons in both cases is approximately
\begin{equation}
V\left(\varphi_{b}^{max}\right)=-\frac{1}{4}\lambda_{-}\left(\frac{32}{\lambda_{-}}\left|\varphi_{f}\right|\right)^{4}=\left(\frac{64}{\lambda_{-}}\right)^{3}\varphi_{f}^{4}\label{eq:71}
\end{equation}
in agreement with \cite{MRS2}. As we have checked, the results obtained
above reproduce not only the parametric dependence but also the numerical
coefficients for the corresponding asymptotic formulas in case of
an exactly solvable potential, as considered in \cite{MRS1}. Moreover,
we have verified that this is also true for the deep tunnelling in
the case of exactly solvable potentials which are constructed piecewise
from quartic and quadratic potentials in three and four dimensions.

\section{Conclusions}

The only known case in which the probability of decay of the false
vacuum can be estimated for general potentials is the case in which
the depth of the true vacuum is much smaller than the height of the
potential barrier separating the false and true vacuums. In such a
case, the thin-wall approximation for instantons applies \cite{Coleman}.
In this work, we have considered the reverse case, i.e., we have assumed
that the depth of the true vacuum (or tunnelling at unbounded potential)
significantly exceeds the height of the barrier. We found that in this
case the corresponding instantons describing the tunnelling are dominated
mainly by the friction term in equation (\ref{eq:2}) and the resulting
bubbles of a true vacuum always have thick walls. This simplifies
the problem considerably and allows us to replace the non-autonomous
 equation (\ref{eq:2}) by the autonomous completely integrable equation (\ref{eq:34}),
which is a good approximation for the original exact equation when
the depth of tunnelling $\left|V_{b}\right|$ significantly exceeds
the height of the barrier $V_{bar}$. As a result, we were able to
derive the general simple formulas (\ref{eq:55}) and (\ref{eq:56})
for arbitrary potentials in any number of dimensions. We applied the
general results to some exactly solvable potentials and showed that
our simple thick-wall approximation exactly reproduces the leading
order asymptotic, including the numerical coefficients, for the deep
tunnellings.

\bigskip{}

\bigskip{}

\textbf{Acknowledgments}

\bigskip{}

The work of V. M. was supported by the Germany Excellence Strategy---EXC-2111---Grant
No. 39081486.

A. S. would like express a gratitude to the Hebrew University of Jerusalem 
for the support and hospitality during his visit. 
The work of A. S. was also supported in part by RFBR grant No. 20-02-00411.


\bigskip{}


\end{titlepage} 
\end{document}